\documentclass[
 reprint,
 amsmath,amssymb,
 aps,
prb,
]{revtex4-2}
\usepackage[caption=false]{subfig}
\usepackage{graphicx}
\usepackage{dcolumn}
\usepackage{bm}
\usepackage[colorlinks,citecolor=blue,filecolor=blue,linkcolor=blue,urlcolor=blue]{hyperref}
\usepackage{xcolor}

\begin{document}

\preprint{APS/123-QED}

\title{Ground state qubit-entanglement of frustrated transverse field models on square lattice.}
\title{Ground state many-body quantum entanglement of a frustrated transverse field Ising model on square lattice.}

\title{Ground state many-body quantum entanglement of frustrated transverse field models on square lattice.}

\author{Vikas Vijigiri}
 \altaffiliation[vikasvikki@iopb.res.in]{}\affiliation{Institute of Physics, Bhubaneswar-751005, Odisha, India}
 \affiliation{Homi Bhabha National Institute, Mumbai - 400 094, Maharashtra, India}
\author{K.G. Paulson}\altaffiliation[paulson.kg@iobp.res.in]{}\affiliation{Institute of Physics, Bhubaneswar-751005, Odisha, India}
\affiliation{Homi Bhabha National Institute, Mumbai - 400 094, Maharashtra, India}
\author{Saptarshi Mandal}\altaffiliation[saptarshi@iobp.res.in]{}\affiliation{Institute of Physics, Bhubaneswar-751005, Odisha, India}
\affiliation{Homi Bhabha National Institute, Mumbai - 400 094, Maharashtra, India}
\author{A. M. Jayannavar }\altaffiliation[jayan@iobp.res.in]{}\affiliation{Institute of Physics, Bhubaneswar-751005, Odisha, India}
\affiliation{Homi Bhabha National Institute, Mumbai - 400 094, Maharashtra, India}

 \date{\today}

\begin{abstract}
 We study the ground state (GS) many-body quantum entanglement of two different transverse field models on a quasi-2D square lattice relevant to a Hydrogen-bonded crystal, i.e, squaric acid. We measure the genuine multipartite qubit-entanglement ($C_{\text{GME}}(\psi)$) of the ground state of very generic models with all the possible cases of exchange couplings considered under defect free and one lattice site defect conditions. Our results show that creation, decay of multipartite entanglement occur for different combinations of coupling strength. When frustration is maximum the system exhibits a peak in concurrence after a gradual increase from disentangled state at zero field followed by an asymptotic decay at large fields. In contrast, for a marginally frustrated (degenerate) case though the concurrence shows a peak, the entanglement is non-zero and large at zero fields. Our results discuss the sensitivity of the qubit-entanglement with underlying GS of varying degree of degeneracy. We conclude that despite of their similarities in ground state properties, yet we see a difference in the degree of entanglement between the two models. We conjecture this result could be due to the difference in the amount of degeneracy and the quantum ground states of both Hamiltonians that could dictate the results even in the thermodynamic limit.
\end{abstract}

\maketitle


\section{Introduction}
It is a well-known fact that entanglement is one of the important features of quantum states that can be availed as a resource for implementing many information and computation protocols~\cite{ekert1991,bennett1992,bennett1993} successfully, that are otherwise classically impossible. On the other hand, as we know, entanglement is also considered as a non-critical resource to implement many non-local~\cite{dakic2012,toth2018,paulson2019} quantum protocols. This point towards the need for complete characterizations of quantum correlations of state space which is non-trivial from both the operational and fundamental perspectives. 
The history of manipulating spin degrees of freedom of a physical system for physical realization of qubits and for implementing  information protocols dates long back. The multipartite quantum coherence and correlations of spin systems are an inevitable resource for quantum information and computation tasks. The connection between quantum correlations and quantum matter is being investigated actively, as it governs many physical phenomena of fundamental interest~\cite{dillenschneider2008,zeng2019}. In this regard, one of the current research interest is to establish a strong connection between quantum correlation and quantum phase transition phenomenon in condensed matter physics~\cite{zeng2019quantum}.\\
\indent Recent studies along these lines have revealed interesting results regarding the topological entanglement entropy~\cite{kitaev2006} from an information-theoretic viewpoint where the quantum correlations can be used as a probe to detect the quantum phase transitions~\cite{lou2006,dillenschneider2008,schwandt2009,li2008,jiang2012}, and also characterize the topological order if there exists any. Similarly, the systems with a high degree of frustration are also of special interest in the field of condensed matter as they exhibit spin-liquid behavior in many cases~\cite{savary2015,wang2015}. We know, quantum spin-liquids are prototypical examples of ground states with non-zero many-body entanglement. Their highly entangled nature imbues quantum spin liquids with unique physical aspects, such as non-local excitations with topological properties, and more~\cite{shannon2012}. Similar to the effects of frustration in magnetic materials, Hydrogen-bonded (H-bond) systems are another class of materials that may experimentally mimic the spin systems (spin-ice rules (IR))~\cite{chern2014,vikas2020,vikas2018,ishizuka2011quantum}. Not only that, they could also be served as an alternative in realizing other quantum effects coming from the zero-point motion of protons, for example, intermolecular charge transfers, localization-delocalization, and etc. It may be noted that many of the earlier studies on H-bond systems focused on the dielectric properties arising from the (anti)ferroelectricity in them. Nevertheless, there are also few works focusing on the aspects of coherent quantum tunneling or the concerted proton motion in these systems from the quantum entanglement perspective. For instance, in a hexagonal ring~\cite{pusuluk2019,benton2012}, it was found that the estimated pairwise correlation function showed the essence of individual quantum tunneling and that in the long-time limit this study of individual proton events is enough to capture the behavior of coherent mechanism.  \\
\indent Deviating from the existing trends, the motivation of our current work is to understand the intricate connection between the degeneracy coming from the strong proton-proton correlations and the quantum entanglement. In other words, we are interested in quantifying a relation between the entropy arising from the proton's several ordering with that of the multipartite proton-proton (qubit) entanglement. We study two different models that have a varying degree of degeneracies \cite{vikas2018,ishizuka2011quantum}. We estimate the multi-partite entanglement for a finite-size spin system, and infer a logical extension of our study in the thermodynamic limit.  We draw our motivation to study the experimentally abundant proton (spin) system, i.e, squaric acid ($\text{H}_2\text{SQ}$) with ice-rules seen as frustration~\cite{ramirez1999}. \newline
\indent Squaric acid ($\text{H}_2\text{C}_4\text{O}_4$) is a network of $\text{C}_4\text{O}_4$ molecules arranged in a quasi-2D  lattice  with protons being in the intermediate position  of every two Oxygen atoms (see Fig.~\ref{hamiltonian}). We use pseudospin modeling of the proton system with protons being represented as pseudospin-1/2 operators of an equivalent quantum spin system. The two models that we analyze our results on quantum entanglement are: 1) A frustrated transverse field Ising model (TFIM) (Model A) and 2) a gauge-invariant extended Ising lattice gauge theory in (2+1) dimensions (Model B). Both the models have shown to be replicating to some extent the experimental findings in $\text{H}_2\text{SQ}$ where the implications of frustration are quite significant~\cite{chern2014,vikas2018,ishizuka2011quantum}. 
There are earlier studies with a finite system size that scale arguably well to understand its qualitative behavior in the $N \rightarrow \infty$ limit and we expect the same for our present study also.\\
\indent Here, we calculate a measure of general multipartite entanglement, concurrence ($C_\text{GME}$) of the ground state as a function of external transverse Zeeman field $K_x$, and the intramolecular strength. We employ the method of exact diagonalization (ED) to obtain the exact eigenstates and then calculate the density matrix.  
We note that both the models have shown to have similar ground states (classical) but different degeneracies~\cite{vikas2018,ishizuka2011quantum}. All the possible cases in both the models are categorized based on their classical ground states and respective degeneracies for a 4-qubit system (plaquette). Later, the qualitative behavior of concurrence is analyzed based on this categorization. Finally, the contrasting differences in the amount of entanglement between the two models are discussed, and a conclusion is made about the sensitivity of qubit-entanglement with the underlying quantum ground states in the thermodynamic limit.\\
\indent This paper is organized as follows. In Sec~\ref{sec2}, we introduce models, and methods of the study followed by the quantification of multipartite entanglement in Sec~\ref{sec3}. After introducing the pseudo-spin model in Sec~\ref{sec3}, we review and give the details of the ground state of Hamiltonian of a single plaquette in Sec~\ref{Gsection}. The results then obtained from the ED analysis of concurrence are given in Sec~\ref{staticcoupling} for various selected coupling strengths for both the models. Similarly, in Sec~\ref{dynamiccoupling}, we discuss the same for the case covering a larger phase-space of $J_1, J_2$ for Model A ($J_0$, $J_1$ for Model B). In the end, we summarize and discuss our results in Sec~\ref{summary}.
\section{ The Models  and the Hamiltonian}
\label{sec2}
As mentioned before, our analyses are based on pseudo-spin formalism, i.e, the strongly correlated $\text{H}_2\text{SQ}$ proton system is effectively represented by pseudo-spins described by a suitable quantum spin-1/2 model. Two different models (which we refered as Model A and  Model B) that were effectively used to describe the ice-rule physics in $\text{H}_2\text{SQ}$ crystal are considered below. It may be noted that both the models were studied in view of the phase transitional aspects of $\text{H}_2\text{SQ}$ with a restriction on Hamiltonian projecting on the states that satisfy ice-rules (IR). That is to say, only the parameter regimes that constrains to satisfy the ice-rules are considered. However, as we are interested in quantum correlation properties, we
consider all possible scenarios with no restriction on the exchange couplings. We note that both the models are based on the mechanism of describing the order-disorder phenomenon in $\text{H}_2\text{SQ}$. The Model A is given by the transverse-field Ising model with competing interactions ($J_1$ ,$J_2$, $J_3$) giving rise to certain degree of frustration depending on the nature of the couplings. Similarly, the Model B is given by the Ising lattice gauge theory (2+1) with degeneracies playing the role of frustration as in Model A. Since we consider qubit (spin) dimensions up to a size of 12, it is therefore convenient to represent the Hamiltonian as a sum of elementary plaquette interactions with each plaquette being occupied by a single molecular unit $\text{C}_4\text{O}_4$. Unlike the Model A (where the IR states are obtained as a consequences of frustration arising from both the couplings, i.e, $J_1$ and $J_2$), the IR states are restrained directly by the Ising gauge term ($J_0$) along with the antiferromagnetic Ising interaction term ($J_1$). In the parameter regimes that we consider, it is shown that both the models closely realize a quantum paralelectric (liquid-like) phase in the quantum Monte Carlo simulations~\cite{vikas2020,chern2014,ishizuka2011quantum}. Therefore, it could be interesting to observe the subtle differences among the two, especially from the ground state quantum entanglement perspective of two completely different Hamiltonians. 
Although both of them depict the same system, we note that the spins in the Model A~\ref{model1} are located on the lattice sites in contrast to Model B~\ref{model1} where they are located on the links (see Fig.~\ref{hamiltonian}). Note that, with periodic boundary conditions the number of spins in Model B is equal to twice the number of spins in Model B.
\subsection{Model A: Frustrated Transverse Field Ising Model}
\label{model1}
\begin{figure}[t!]
\hspace{3.0cm}
\includegraphics[width=8.5cm,height=6.3cm]{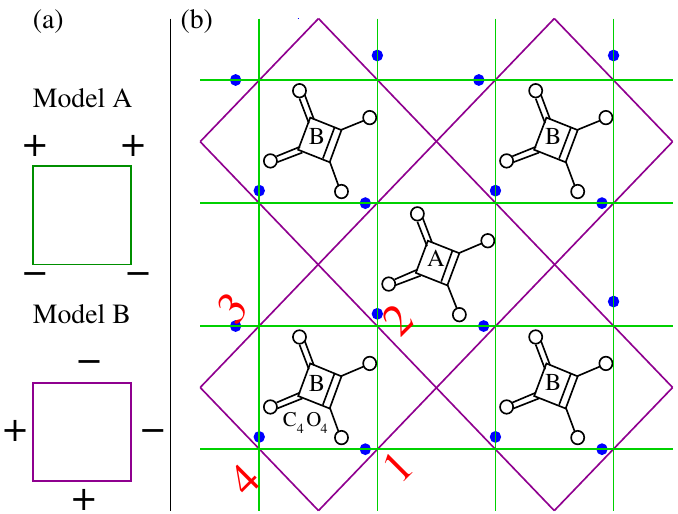}%
\caption{(Color online) (b) Schematic 2D square lattice showing the leading interactions $J_{1,2,3}$ ($J_{0,1,2}$) for Model A (B). The small blue dots represent the qubits, i.e, the protons (or pseudospins). (a) Shows the arrangement of spins for Model A and similarly, lower panel (purple) corresponds to Model B. The details of the interactions are given in Sec~\ref{sec2}}
\label{hamiltonian}
\end{figure}

The TFIM models have been pivotal in describing the simpler H-bonded systems since earlier times, for example they are extensively used to describe the ice-rule physics in $\text{KH}_2\text{PO}_4$ (KDP)~\cite{matsushita1982note}, however, the number of competing interactions and the nature of them are completely system dependent. Here, we consider one such TFIM model with competing interactions, and the terms that we account for has already been briefly discussed before \cite{ishizuka2011quantum}, and is given below,
\begin{eqnarray}
\label{Hamiltonian}
H &=& J_1\sum_{\langle ij \rangle} \sigma^z_i\sigma^z_j+ J_2\sum_{[ ij ]} \sigma^z_i\sigma^z_j - J_3\sum_{( ij )} \sigma^z_i\sigma^z_j + \nonumber \\
&& K_x\sum_{i}\sigma_{i}^{x} ,
\label{hmt1}
\end{eqnarray} 
where $\sigma^{x,y,z}_i$ are the Pauli matrices representing the pseudo-spin  operator, $\sigma_i$, at each site $i$ (see Fig.~\ref{hamiltonian}(a)). The parentheses $\langle\rangle$, $[]$, $()$ denote the  pairs of sites corresponding to $J_1$, $J_2$, and $J_3$ as nearest, next-nearest (N) and next next-nearest (NN) neighbor interactions respectively, as shown in Fig.~\ref{hamiltonian}. While the external field $K_x$ is the on-site transverse magnetic field. Note that this external field can be experimentally related to pressure. Since, the effect of pressure is to reduce the coupling constant thereby increasing the rate of tunneling motion of protons (or more quantum fluctuations for pseudo-spins). The nature of coupling constants we consider encloses all possible cases of couplings, i.e, $J_i \hspace{2mm} \varepsilon \hspace{2mm} \mathcal{R}$ ($i\equiv \{1,2,3\}$). In order to see the frustration arising because of $J_2$, the minimum qubits required are $N_Q=4$, i.e, one unit plaquette. Similarly, for $J_3$ it is $N_Q=12$ with 4 plaquettes along $x$ and 3 plaquettes in $y$. 
\graphicspath{{Frustrated TFIM/}}
\\
\subsection{Model B: Ising Lattice Gauge Model in (2+1) dimension}
\label{model2}
\begin{figure}[t!]
\hspace{3.0cm}
\includegraphics[width=8cm,height=7cm]{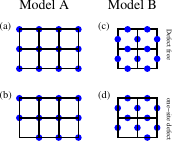}%
\caption{(Color online) Illustration of the 12-site clusters for both the models. (a) and (b) correspond to the case of cluster with and without defect respectively for Model A. Similarly, (c) and (d) correspond to same for Model B.} 
\label{cluster}
\end{figure}

Similarly, the other model that is extensively used here is the extended version of Ising lattice Gauge theory in (2+1) dimensions, the motivation for such modeling historically comes from the vertex-based models~\cite{stilck1981vertex}. The gauge term may be understood to be originated primarily because of the interaction mediated by the $\pi$-electrons in $\text{C}_4\text{O}_4$ by the rearrangement of $\pi$-bonds. Further details describing the possible mechanism, and the effective vertex model can be found in Ref.~\cite{stilck1981vertex}. Nevertheless, the Hamiltonian is given as,  
\begin{equation}
    H=-J_{0}\sum_{\square}A_{\square}+J_{1}\sum_{\square}B_{\square}-J_{2}\sum_{\langle AB\rangle}\vec{P}_{A}  \cdot \vec{P}_{B}-K_x\sum_{i}\sigma_{i}^{x},
    \label{hmt2}
    \end{equation}
where, $A_{\square}=\sigma_{1}^{z}\sigma_{2}^{z}\sigma_{3}^{z}\sigma_{4}^{z}$, the indices $1,2,3,4$  represent the four spins located on the edges of a plaquette, while $B_{\square}=\sigma_{1}^{z}\sigma_{3}^{z}+\sigma_{2}^{z}\sigma_{4}^{z}$ with the indices $1,3$ conveniently located diagonally opposite to each other within a given plaquette. Here, $P_{A,B}^{x}=\pm(\sigma_{1}^{z}+\sigma_{4}^{z}-\sigma_{2}^{z}-\sigma_{3}^{z})$ and $P_{A,B}^{y}=\pm(\sigma_{1}^{z}-\sigma_{2}^{z}-\sigma_{3}^{z}+\sigma_{4}^{z})$ are the dipole-moment vectors for $A,~B$ sub plaquettes. Here, with PBC the no. of plaquettes are 2, 2 (12-site cluster) along x and y directions as depicted in Fig.~\ref{cluster}. However, we note that $J_2$ has terms with next to  next nearest neighbour interactions for which the lattice size required is a minimum of 4$\times$4. However, we are interested in the physics of frustration where $J_2=0$~\cite{vikas2018,chern2014} because in the presence of $J_2$, the system develops a long range ordered states.
So, in the present work we consider the entanglement dynamics under different parametric conditions with different qubit sizes $N_Q\in\{4, 12\}$ corresponding to 1 and 4 plaquettes. .
\\
\begin{table}[tp]
    \centering
    \caption{Nature of GS under various parameter regimes}
    \label{tbl:crossref-label}
    \begin{tabular}{l d d d d @{\extracolsep{12pt}}d d d d l}
        \toprule
        &    & \multicolumn{2}{c}{$J_1>0$} & & \multicolumn{3}{c}{$J_1<0$} \\ \\
        $J_2 <0$ &  &  \multicolumn{2}{c}{$\text{Partial}$} & & \multicolumn{3}{c}{$\text{FM}$} \\ \\
        $J_2 >0$  &  &     \multicolumn{2}{c}{$\text{Frustrated}$} & & \multicolumn{3}{c}{$\text{Partial}$} \\
        &&&&& \\
    \end{tabular}
\end{table}
\section{Multipartite quantum correlations}
\label{sec3}
Among all the various measures that are employed to measure the quantum correlations of a multipartite state, quantum entanglement is a significant one among them. An n-partite pure state $\mid\psi\rangle$ $\in$ $H_1\otimes H_2\otimes....\otimes H_{n}$ is called bi-separable if it can be written as a product of two pure state as $\psi=\mid\psi_{A}\rangle\otimes\mid\psi_{B}\rangle$, where $\mid\psi_A\rangle\in H_{A}=H_{i_{1}}\otimes....\otimes H_{i_{j}}$ and $\mid\psi_A\rangle\in H_{B}=H_{i_{i_{j+1}}}\otimes....\otimes H_{i_{n}}$. If an n-partite quantum state in not bi-separable then it is called as n-partite entangled state. so, genuine multipartite entanglement $(\text{C}_{\text{GME}})$ of an n-partite pure state~\cite{ma2011} is given as
\begin{equation}
    \text{C}_{\text{GME}}(\mid\psi\rangle)=\min_{\gamma_{i}\in\gamma}\sqrt{2(1-\text{Tr}(\rho_{A_{\gamma_{i}}}^2))},
    \label{cgme}
\end{equation}
where $\gamma={\gamma_{i}}$ is the set of all possible bipartitions. We use Eq.~\eqref{cgme} to calculate the multipartite ground state entanglement of the system under various parametric conditions, and the concurrence versus external field obtained along various contours are discussed in detail in the next sections.
\begin{figure}[t!]
\includegraphics[width=1.0\linewidth]{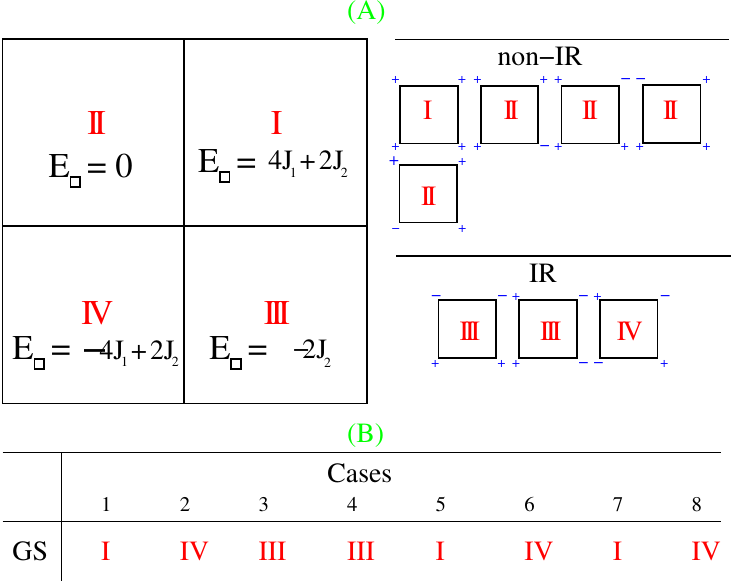}
\includegraphics[width=1.0\linewidth]{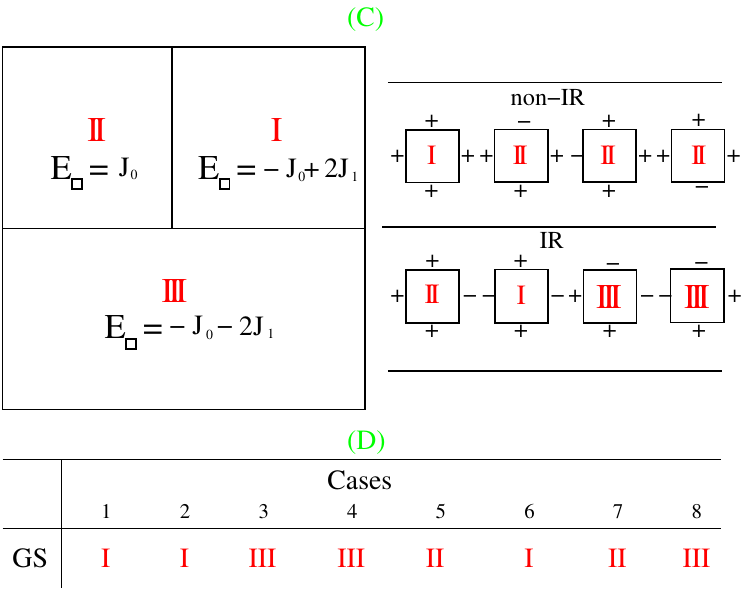}
\caption{(Color online) Schematic presentation of the configurational sets and their respective ground states (GS) energies. (A) GS energy of sets labeled by I, II, III, IV for Model A ((A), (B)).  Similarly, (C) and (D) are analyzed for Model B.}
\label{GS}
\end{figure}
\section{Ground States; Case by Case}
\label{Gsection}
As our primary emphasis is on understanding the role of frustration, we, therefore, limit our discussions to the case where $J_3=0$. Since we know from the earlier studies~\cite{ishizuka2011quantum}, turning on the $J_3$ interaction drives the system to a global ordered antiferroelectric state where the degeneracy is lifted (no frustration) and that is independent of system size. 
Besides, the dynamics of the qubits depend strongly on the nature of interactions, and it becomes important to distinguish the classical ground states that may dictate the amount of frustration and thereby change the degree of entanglement in the quantum version of it. We start with a system of four qubits and extend our discussion based on the arguments provided below to higher lattice dimensions.
\subsection{Model A}
\label{modelA}
\begin{figure}[t!]
\includegraphics[width=1.0\linewidth]{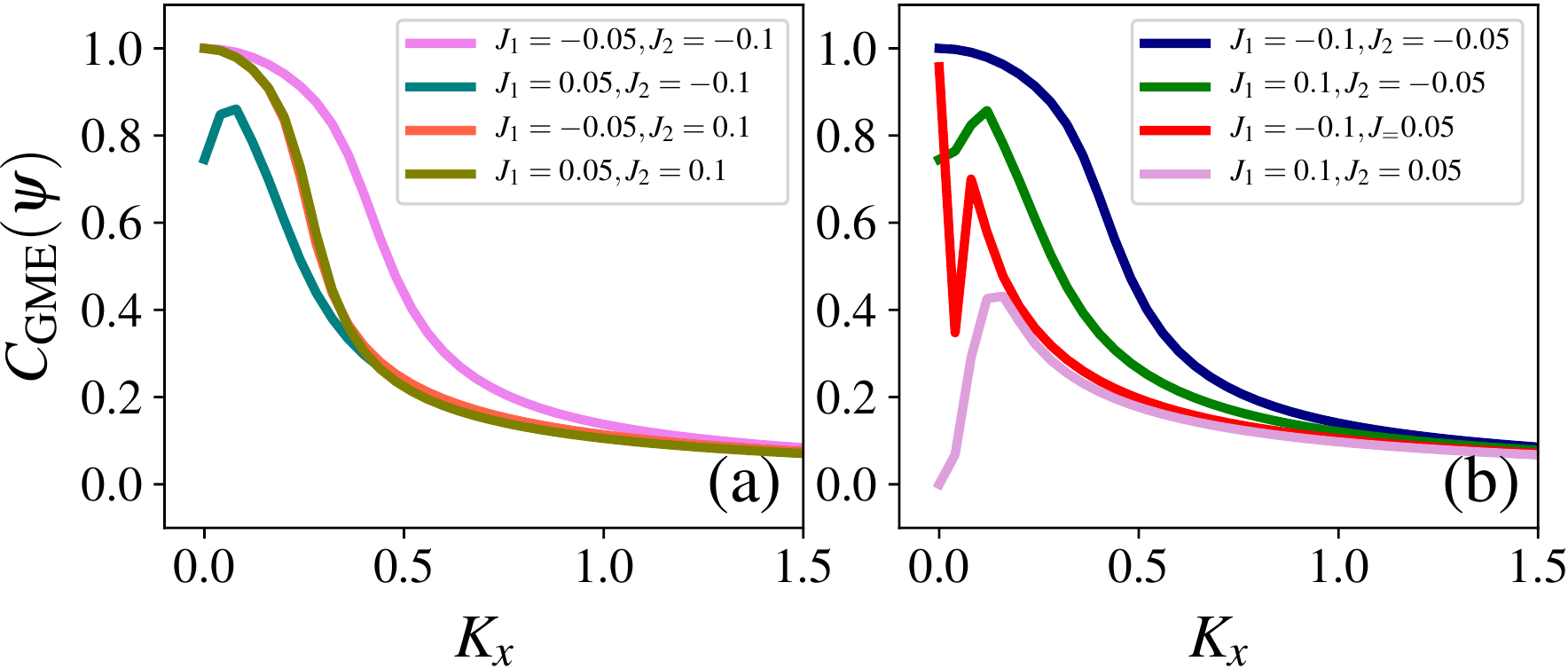}
\includegraphics[width=1.0\linewidth]{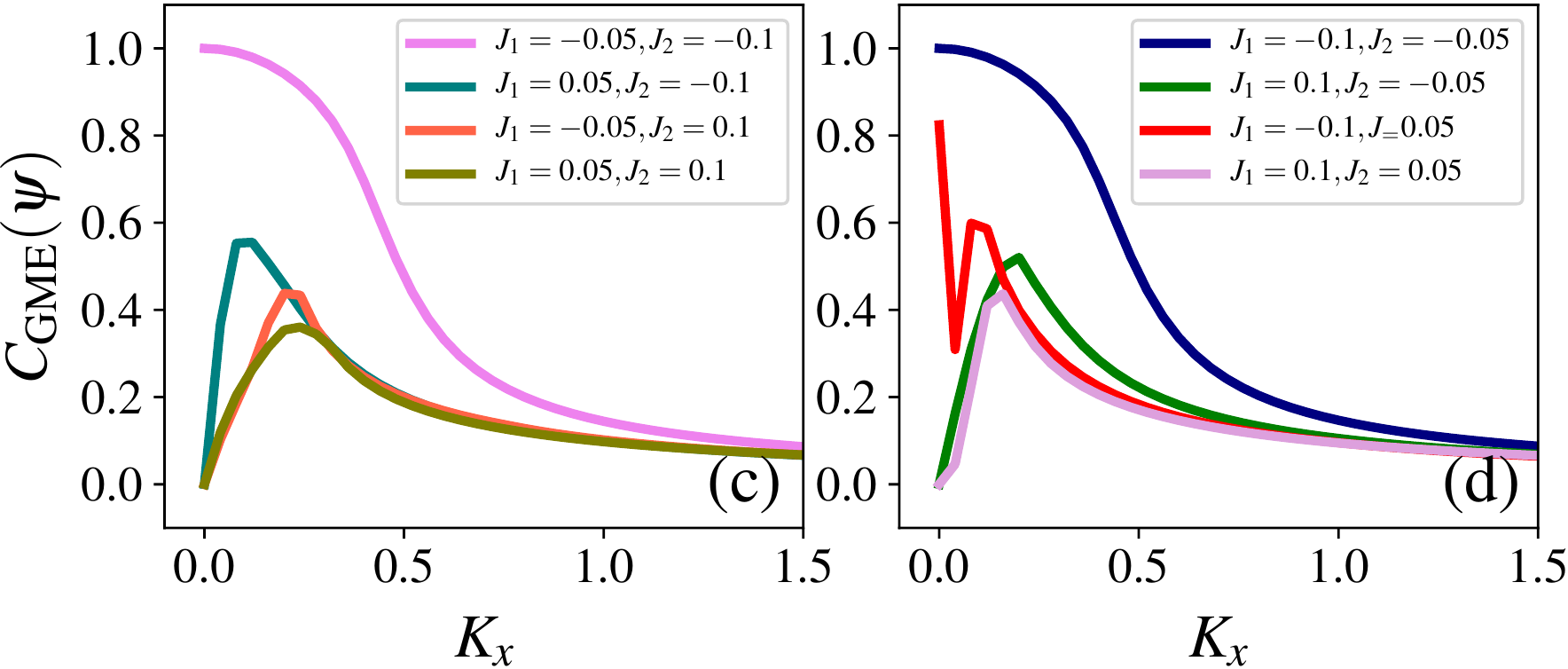}
\caption{(Color online) Plots showing the variation of general multipartite entanglement measure, concurrence ($C_{\text{GME}}(\psi)$) as a function of field strength $K_x$. Here, (a)-(b) are calculated for a 12-site cluster without any defects and the plots (c)-(d) are shown for one lattice site-defect. Periodic boundary condition (PBC) is invoked. All the results here correspond to Model A.}
\label{stat_loc}
\end{figure}

\indent Now, consider the following case of zero-field. All the GS configurations, and their corresponding energies for the Hamiltonian are shown in the schematic in Fig.~\ref{GS}. Where the labels I, II, III, and IV denote the sets of configurations categorized based on their GS energies estimated on a single given plaquette. Here, $J_1$ and $J_2$ both can be either ferromagnetic (FM) or antiferromagnetic (AFM) type. There are thus four sets of combinations. 1) $J_1<0; J_2<0$ 2) $J_1>0; J_2>0$ 3) $J_1<0; J_2>0$ 4) $J_1>0; J_2>0$ and in each set we consider sub-cases where (i) $\mid J_1 \mid < \mid J_2 \mid$, (ii) $\mid J_1 \mid > \mid J_2 \mid$, thus totalling to eight combinations. Noticeably, not every case has the same set of GS configurations shown in sub-Fig.~\ref {GS}. We chart out the ground states in each of the eight cases in the chronological order plotted in Fig.~\ref{stat_loc}, for instance, case 1 (Fig.~\ref{GS}) corresponds to the plot (a) (orange) in Fig.~\ref{stat_loc}, and so on. \\
\indent The configuration set I, II, III, IV have the energy $E_{\square}$ which is given by $4J_1+2J_2$, 0,  $-2J_2$, $-4J_1+2J_2$ respectively with 2, 8, 4, 2 associated degeneracies respectively. For example, consider the case 3~\ref{GS}(B) corresponding to the values of $J_1=-0.05$ and $J_2=0.1$, the GS is given by the configuration set III with energy per plaquette, $E_{\square}=-0.2$. Similarly, for case 2, $J_1=0.05$ and for $J_2=-0.1$, the state III is no more the GS but instead IV with energy per plaquette, $E_{\square}=-0.4$. Likewise, for $J_1=-0.05$ and for $J_2=0.1$ it is again given by the set III ($E_{\square}=-0.2$) and so on. This way all the GS for the respective parameter space were obtained case by case as shown in Fig.~\ref{GS}(B). Furthermore, from Fig.~\ref{GS}(B) it is also clear that only the states corresponding to I, III, IV tend to appear in at least one of the 8 cases. The configuration set II which is a state with larger degeneracy do not appear in any of the cases mentioned. However, we have a state III with twofold degeneracy within a plaquette and that scales exponentially with system size. Therefore, for frustration effects it is natural to look at the case where the set III is the GS, such as case 3 and 4 (see Fig.~\ref{GS}(B)). \\
\indent Note that the globally ordered states corresponding to the Greenberger-Horne-Zeilinger (GHZ) states do contribute to the maximum entanglement, this is possible when the GS has twofold degeneracy, since in Fig.~\ref{GS}(A) the GS with all the spins flipped also correspond to the same energy. For example, the configurational state I is a state with all spin-up in a plaquette (Fig.~\ref{GS}), so the flipped configuration of it, i.e, the state with all spin-down is also a state with the same energy as the configurational state I.\\
\indent We divide the configuration set basically into the ones that satisfy (IR) and don't satisfy (non-IR or ionic defects) ice-rules. The important thing to notice, however, is that due to frustration (IR) in the system, the net contribution to the concurrence doesn't need to come alone from one particular basis state. Quantum correlations also arise from the superposition principle. Nonlocal correlations found in nonseparable quantum superposition states are known as quantum entanglement. Indeed, because of the degeneracy arising due to frustration, the most general ground state is given by the superposition of all the degenerate candidates in their respective basis sets. For example, for set III, the GS is, $\mid \psi_{\text{GS}}\rangle\equiv  \frac{1}{2}(\mid \text{III}_1\rangle + \mid \text{III}_2\rangle + \mid \text{III}_3\rangle + \mid \text{III}_4\rangle)$. And since there are GHZ states $(\frac{1}{\sqrt{2}}[\vert0000\rangle]+\vert1111\rangle)$ involved in some cases, we, therefore, see a case of maximum initial entanglement which may gradually decrease as we increase the field-strength. The GHZ states appear when the degeneracy in the system is only two-fold. \\
\subsection{Model B}
We see that the same set of configurations that featured in the Model A~\ref{GS} also observed to be the classical GS ($K_x=0$) of this case. Here, we extend the discussion of classical GS in Sec.~\ref{Gsection} to the Hamiltonian (Model B) comprising Ising gauge interactions with each gauge copy being independently representing the degenerate configuration. 
Unlike the Model A~\ref{GS}, here there are three sets of configurations with lowest energies, i.e, I, II, III (convention not according to Fig.~\ref{GS}) with corresponding energy per plaquette given as, $E_{\square}=-J_0 + 2J_1$, $J_0$, and $-J_0-2J_1$ respectively. Note that the ice-rule states are the states in the system that are relevant to experiments, and the parameter regimes that have IR states as their lowest energy states are critical to the experimental findings. Nevertheless, in contrast to Model A, the degeneracies here are 4,8,4 respectively for a single given plaquette for set I, II, III. Important difference is that in all cases the configuration sets appear to have degeneracies that do scale with system size which is of an advantage. We now have the basic idea of GS, and the configurations of IR and non-IR states. In the next section, we present the results case by case and see how the results vary differently depending on the nature of the coupling concerned with the underlying GS, and infer the variation of entanglement.
 \begin{figure}[t!]
\includegraphics[width=1.0\linewidth]{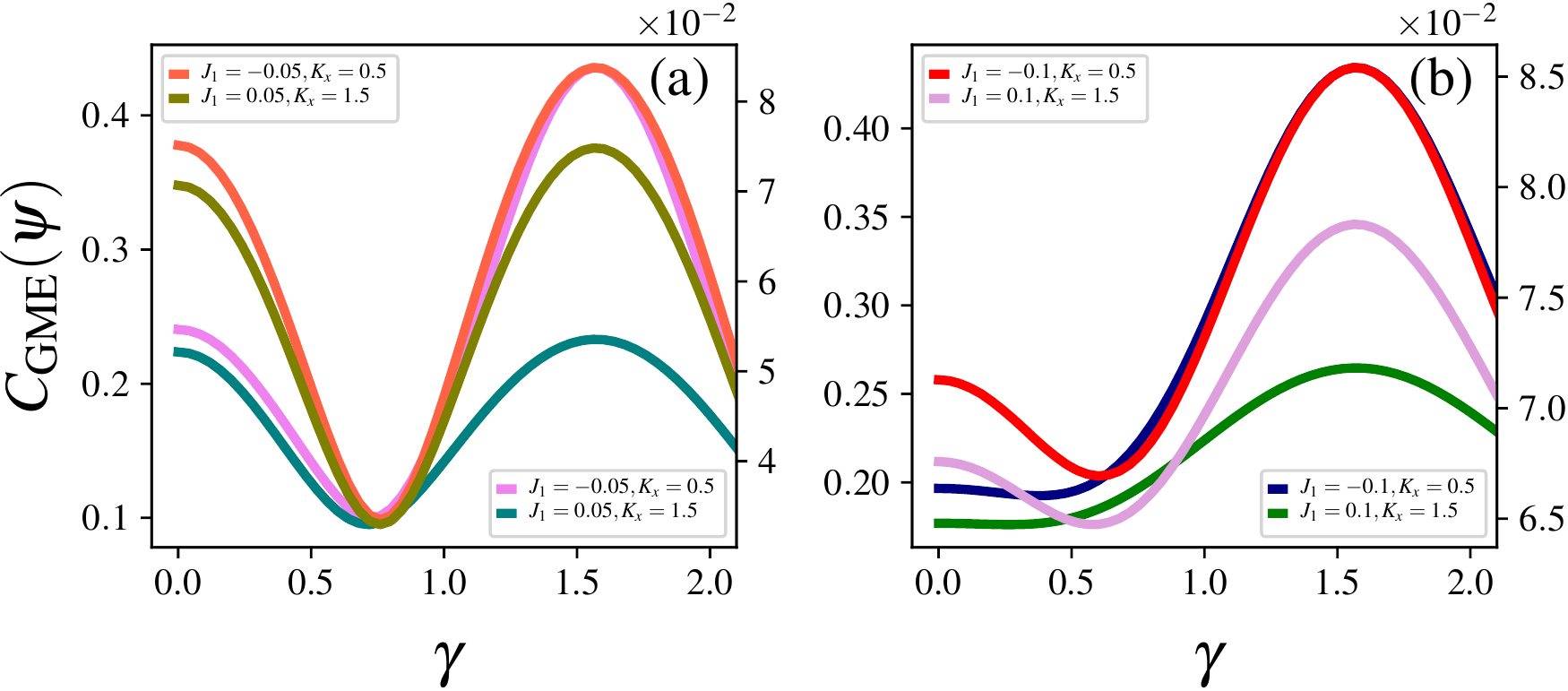}
\includegraphics[width=1.0\linewidth]{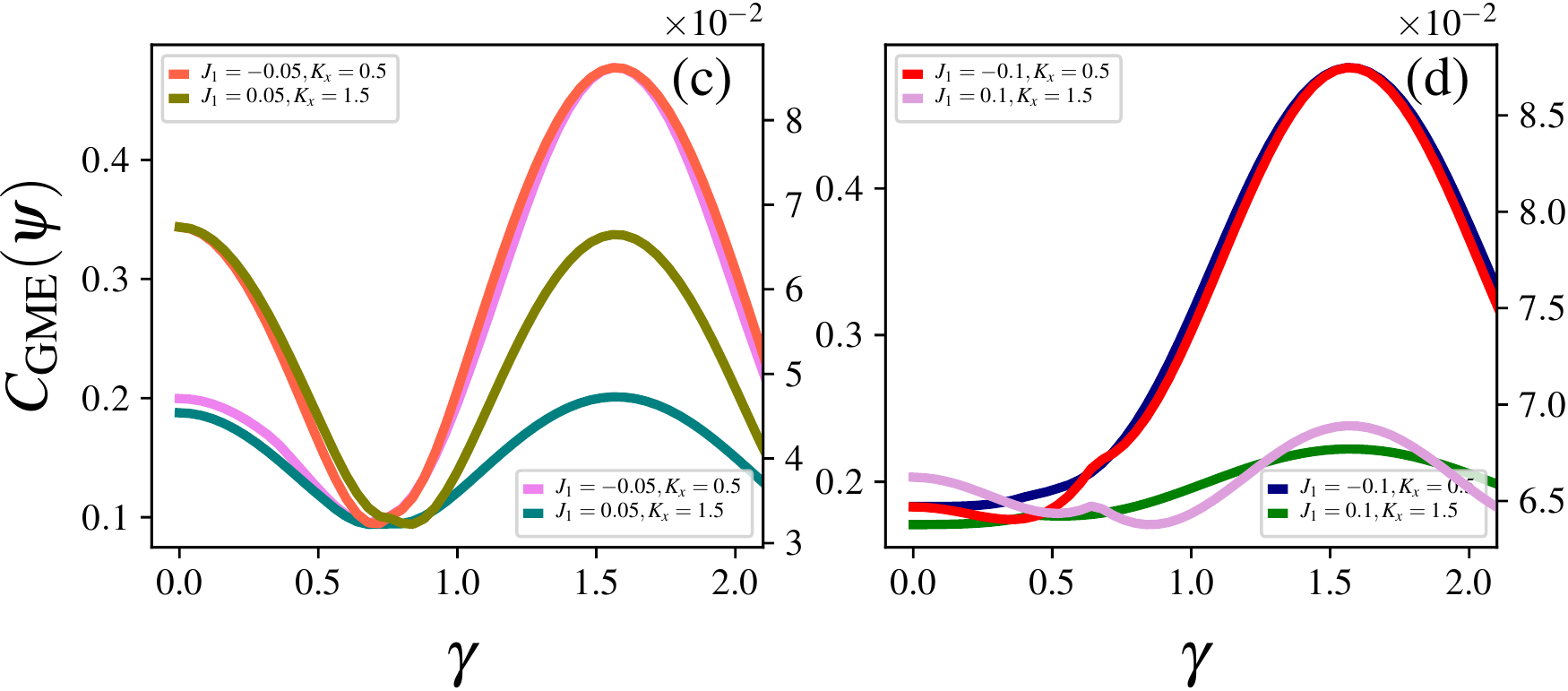}
\caption{(Color online) Concurrence as a function of $\gamma$ is shown for various values of external field, $K_x$ and $\gamma$ (where $J_2(\gamma)=J_2\cos(\omega \gamma)$), where $\omega=0.2$. Here, (a)-(b) are estimated for a 12-site cluster without any Defect, and the plots (c)-(d) are shown for one lattice Defect. We use 12 qubits as the system size with PBC. Here, all the results correspond to Model A.}
\label{dyn_modelA}
\end{figure}
\section{Concurrence versus external field}
\subsection{Model A}
\label{staticcoupling}
\textit{Defect free:} We outline the results mainly obtained for a system of 12-site cluster (see Fig.~\ref{cluster}). In Fig.~\ref{stat_loc}, the variation of concurrence is plotted for  different combinations of coupling strengths. As it is seen, it consists of eight plots with each case corresponding to various combinations of coupling strengths $J_{1}$ and $J_{2}$ in the chronological order that is mentioned in Sec.~\ref{modelA}. From Fig.~\ref{GS}, we obtain that for the case of $J_{1}=-0.05, J_{2}=-0.1$ (Fig.~\ref{GS}(a)), the concurrence at $K_x=0$ is maximum, and is close to 1, further upon increasing the quantum fluctuations induced by the external field, the concurrence gradually decreases asymptotically to 0, as anticipated. Note that the classical GS for this case is the set I (see Fig.~\ref{GS}(B)) which is a GHZ state belonging to the non-IR category.  In contrast to this, the results for $J_1=0.05, J_2=-0.1$ show a gradual increase as the field is increased peaking at around $K_x=0.18$ before gradually decreasing asymptotically as in the previous case. This is a case of the classical ground state having the configurational set IV (see Fig.~\ref{GS}(B)) which is a state satisfying IR. Interestingly, following our analysis of GS in Fig.~\ref{GS}, case 3 and case 4 curves coincide exactly. This is because the underlying GS (set III in Fig.~\ref{GS}(B)) is the same in both cases, as can be seen in Fig.~\ref{GS} where the plot in orange ($J_1=-0.05, J2=0.1$) coinciding with that of olive ($J_1=0.05, J_2=0.1$).  Again, here the set III which is also an IR state with finite-molecular polarization is the underlying GS. Obviously, despite showing qualitative differences in concurrence at low fields, yet the curves almost coincide asymptotically as the field strength is increased where the GS is usually the quantum paraelectric phase. \\
\indent Next, plot (b) in Fig.~\ref{stat_loc} is obtained for the case $\mid J_1 \mid > \mid J_2 \mid$. One immediate difference between the above discussed plot (a) to that of (b) is that the qualitative behavior of concurrence follows a gradual increase with a peak followed by asymptotic decay for larger fields. Here, the plot in blue in Fig.~\ref{stat_loc}(b) corresponds to case 5 and so on of Fig.~\ref{GS}. Again, for the ferromagnetic coupling ($J_1=-0.05, J_2=-0.1$ (case 5)) corresponding to GHZ states has a similar behavior as the case 1. Both the cases 1 and 5 are ferromagnetic and there is no frustration involved. Similarly, case 6 with $J_1=0.1, J_2=-0.05$ shows the same behavior as in the case of case 2 in the plot (a). This is again because the GS being the same (see case 2 and case 6 GS in Fig.~\ref{GS}(B)) in both cases, i.e, set IV. \\
\indent The GS of cases 7 and 8 have set I and IV in contrast to cases 3 and 4 where it is the set III. Clearly, the difference in GS reflects the difference in concurrence variation in plots (a) and (b). However, despite having the same GS, the concurrence values of cases 5, 6 do not match with that of 7, 8 respectively. This discrepancy fundamentally arises from the fact that case 8 is a fully frustrated case with all the bonds including both the coupling $J_1, J_2$ which are not satisfied, while case 6 is not. Indeed, case 6 corresponds to frustration due to $J_2$ bonds only, i.e, only $J_1$ bonds are satisfied. Therefore, as the nature of the coupling $J_1$ is changed from marginally frustrated (case 6, i.e, Fig.~\ref{stat_loc}(b), green) to fully frustrated (case 8, i.e, Fig.~\ref{stat_loc}(b), pink), we, therefore, observe a qualitative difference where the case 8 concurrence starts with zero-entanglement for zero-field and then gradually peaking at 0.2 before again gradually decreasing asymptotically.   \\
\indent \textit{Defect:} This is a case where one site is removed from the 12-site cluster. In Fig.~\ref{stat_loc}(c)-(d), the concurrence is plotted for one site defect as a function of $K_x$. Our analysis proceeds parallel to the fore mentioned discussion. In Fig.~\ref{stat_loc}(c), except the case of ferromagnetic coupling, i.e, case 1 ($J_1=-0.05, J_2=-0.1$) the qualitative behavior of concurrence gets changed compared to the cases of defect free (Fig.~\ref{stat_loc}(a)). Here the non-GHZ states, i.e, non-global ordered frustrated case show a peak behavior similar to the case of fully frustrated case. Intriguingly, when a spin from a lattice site corresponding to an satisfied bond is removed, we see that the system falls back case of frustration. While from the Fig.~\ref{stat_loc}(d) when the bond already unsatisfied is removed there is no qualitative change in the concurrence, see the plot in pink in Fig.~\ref{stat_loc}(d), i.e, corresponding to the case 8. Clearly, this suggests that the cases of fully frustrated and zero frustration cases (GHZ case), the concurrence is insensitive to the site-defects.  Also, regardless of whether the GS is an IR or non-IR state, we see that the concurrence falls to zero in the one site Defect case. It shows that the concurrence is sensitive when in the case of frustration, i.e, the more the frustration the more the effect on concurrence when a defect is put. However, the high field behavior shows no change at all in any of the scenarios regardless of the coupling.\\
\indent Similarly, in plot (d), the concurrence follows same as in the case in plot (a), we note that a similar analysis is valid and can be extended here. The concurrence for plots shown in green, red, pink in Fig.~\ref{stat_loc}(d) show a gradual increase till a certain field strength of approximately $K_x=0.23$ beyond which it starts to decay asymptotically. But, case 8 shows interesting behavior here, we observe that the concurrence curve in the plot (b) and plot (d) i.e, when the system is highly frustrating (pink in Fig.~\ref{stat_loc}(c) and (d)) the concurrence does not change much. It can be understood that the frustration when induced by the less dominated term (here $J_2$), the concurrence shows no to minimal effect under one-site defect. In the next section, we present the results of concurrence varied along the contour covering numerous values of $J_2$.
\begin{figure}[t!]
\includegraphics[width=1.0\linewidth]{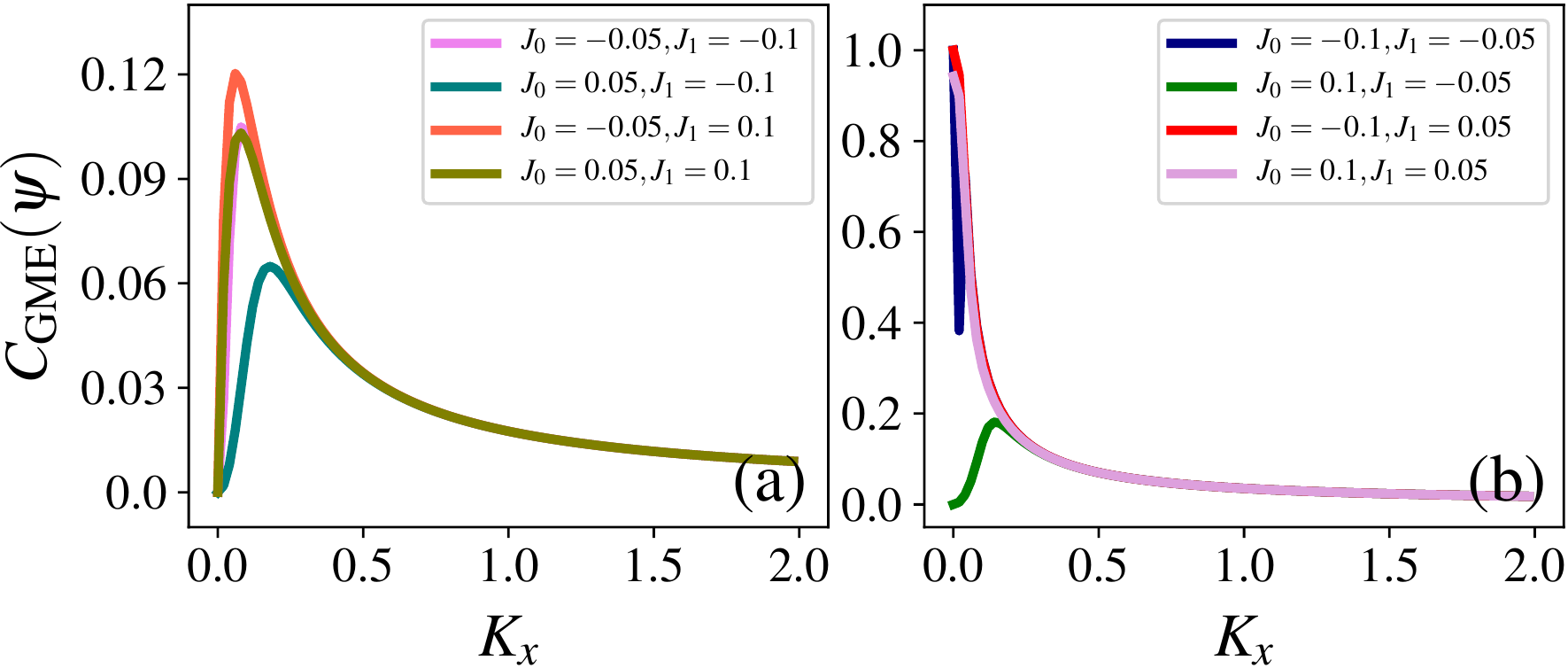}
\includegraphics[width=1.0\linewidth]{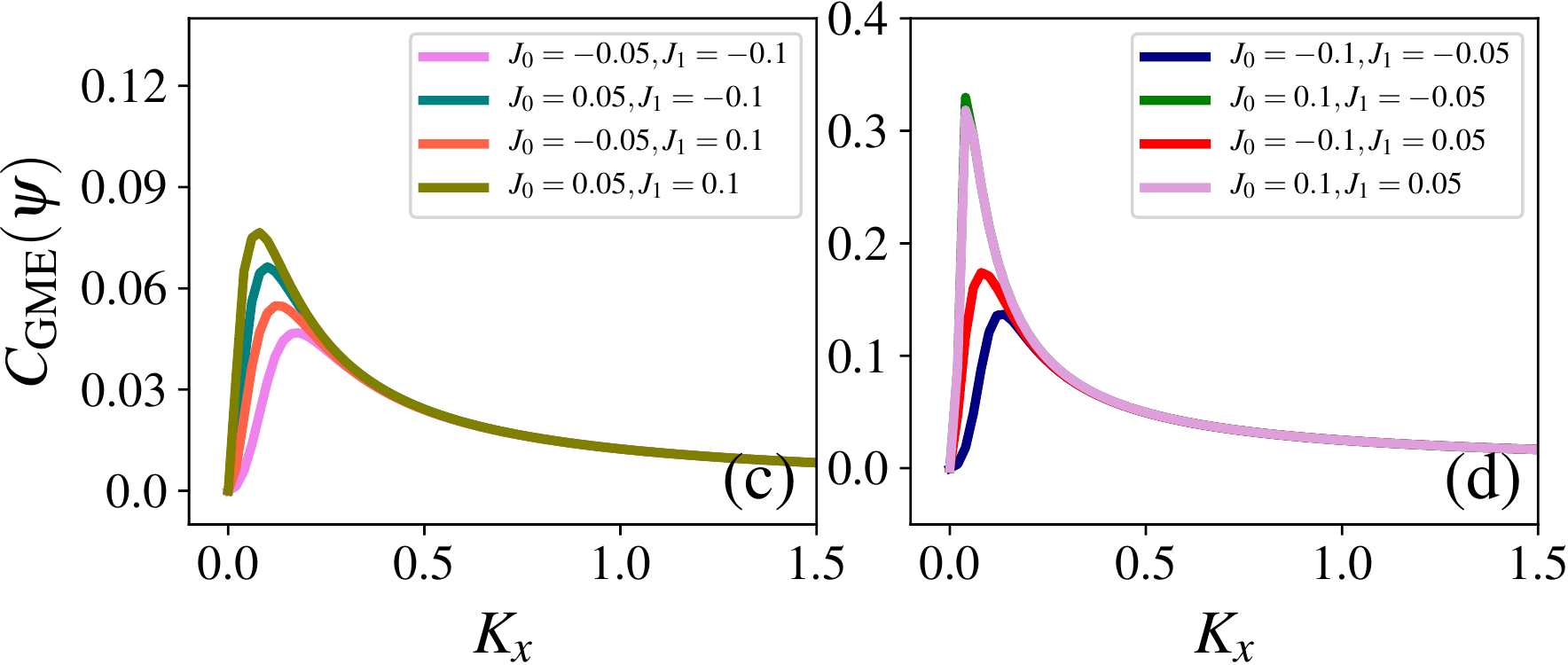}
\caption{(Color online) Variation of concurrence as a function of external field $K_x$ calculated for various parameter regimes. The plots are obtained for Model B, and the system size used is $N_Q=12$ with PBC. The plots (a) and (b) correspond to defect free and similarly (c), (d) to one-site defect respectively.}
\label{H2SQ_stat_loc}
\end{figure}
\begin{figure}[t!]
\includegraphics[width=1.0\linewidth]{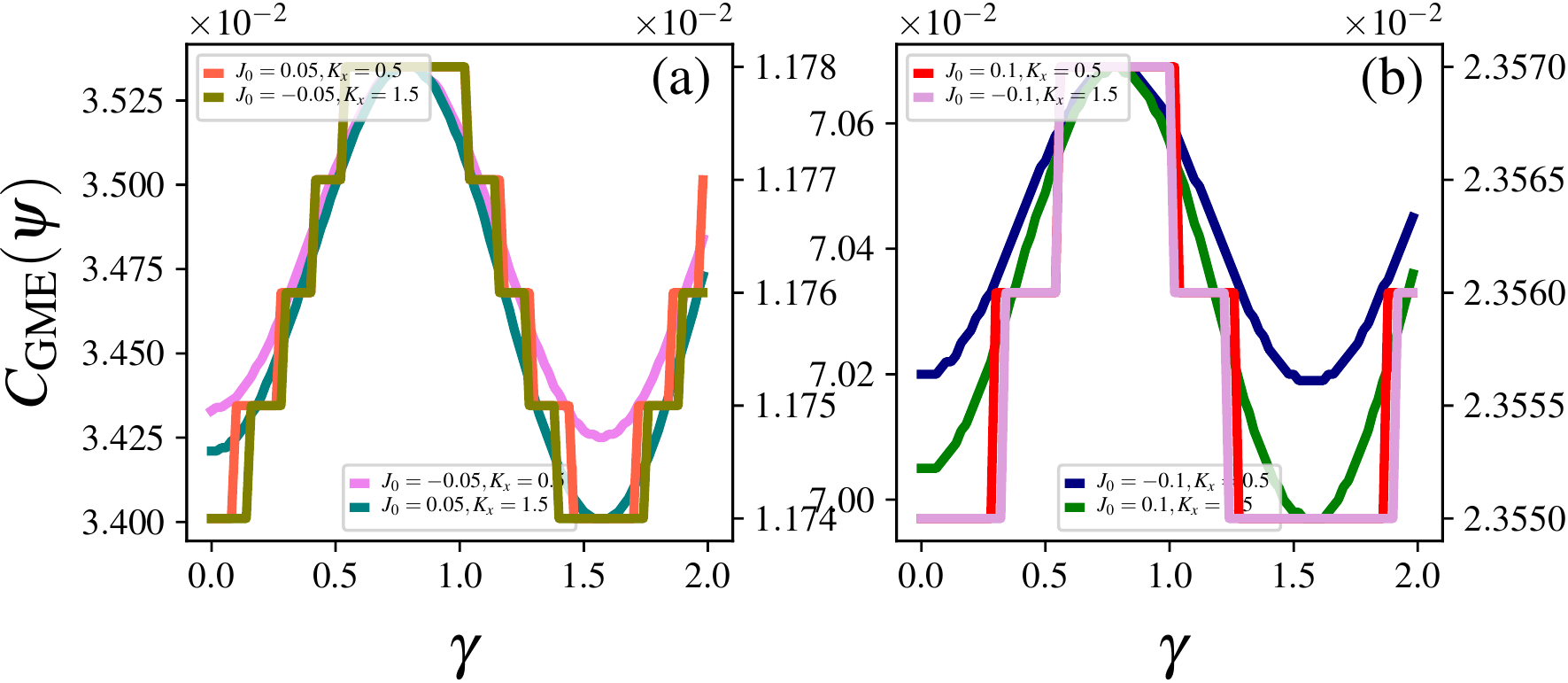}
\includegraphics[width=1.0\linewidth]{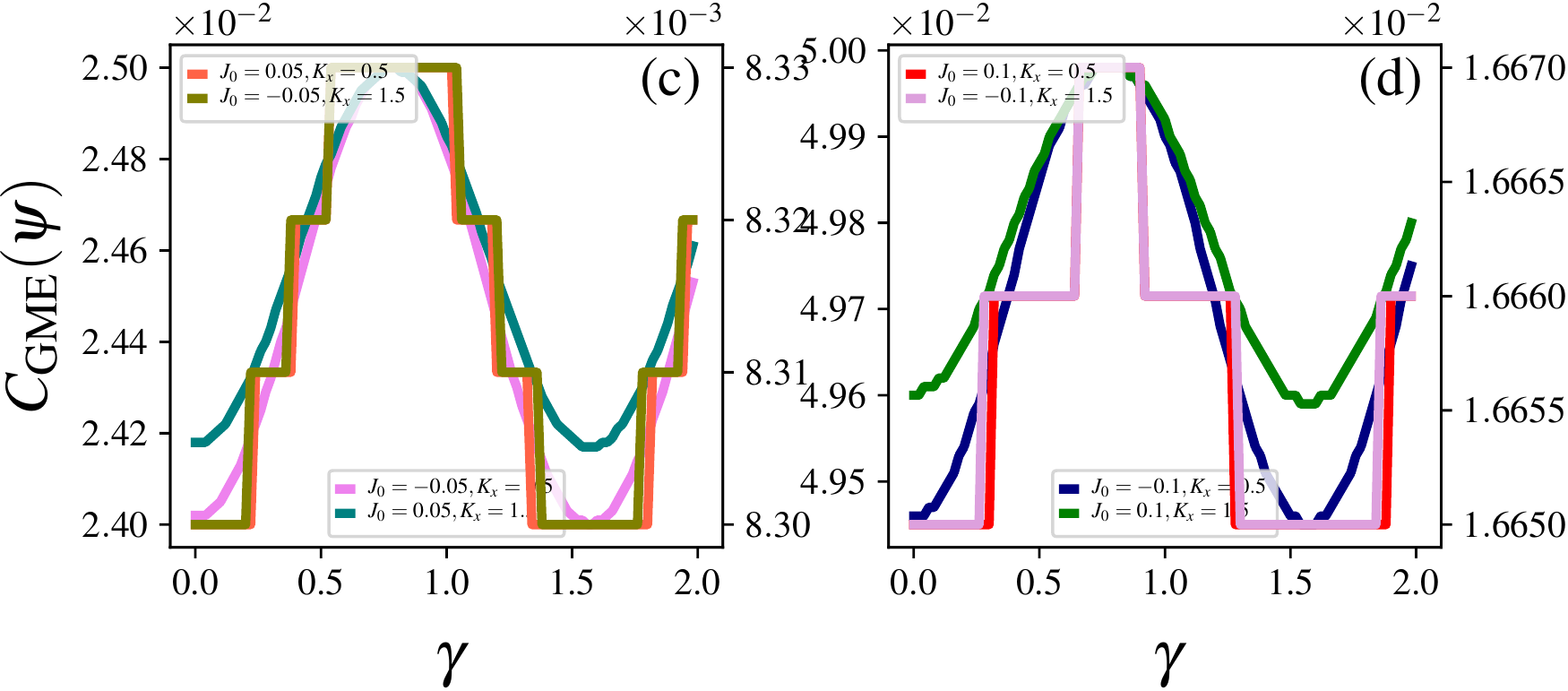}
\caption{(Color online) Variation of general multipartite entanglement measure, concurrence($C_{\text{GME}}(\psi)$) as a function of $\gamma$. Here, (a)-(d) are calculated for a 12-site cluster without any site Defect, and the plots (e)-(h) are shown for one lattice Defect. All the plots here correspond to Model B.}
\label{H2SQ_dyn_loc}
\end{figure}
\subsection{Model B}
\textit{Defect free:} For the Hamiltonian given by the Eq.~\eqref{hmt2} describing the GS properties of $\text{H}_2\text{SQ}$ from a gauge-invariant theoretic perspective, we plot the concurrence (see Fig.~\ref{H2SQ_stat_loc}) obtained for various parameter regimes including the ones that host IR states. The primary difference between Model A with that of B is that there are only three possible configurations with three different GS energies per plaquette. The three configurational sets are labeled by I, II, III, with GS energies, $E_{\square}=-J_0+2J_1, J_0, -J_0 - 2J_1$ respectively.  In Fig.~\ref{GS}(C) and (D), we have presented the classical GS, the corresponding configurations, and their energies, analogous to Model A (Fig.~\ref{GS}(A) and (B)). \\
\indent From Fig.~\ref{H2SQ_stat_loc}, we see that all the cases follow the same qualitative behavior that the concurrence for low values of $K_x$ gradually increases, and then peaking around $K_x=0.25$ before slowly decaying asymptotically. Moreover, the concurrence peak value (0.12) is very small compared to Model A for cases 1-4. We verify that the coupling strength needs to be higher for Model B to have the same degree of entanglement as compared to Model A. In other words, the concurrence peak directly depends on the strength of $J_0$. Our analysis also verifies that a minimum of an order of magnitude 10 is required for such comparison. Furthermore, in plot (b) of Fig.~\ref{H2SQ_stat_loc}, the concurrence varies differently for cases 5, 7, and 8. The concurrence at $K_x=0$ is non-zero, indeed for cases 7, 8 the concurrence is maximum, i.e, close to 1. Consider case 4 in Fig.~\ref{H2SQ_stat_loc} (a) and case 8 in Fig.~\ref{H2SQ_stat_loc} (b), we see that the GS for both the cases is set III (see Fig.~\ref{GS}(D)), however, the qualitative behavior is different for both of them. This can be understood in a way  that when $J_1 \gg J_0$, the system can be decoupled to 1D molecular chains along interpenetrating orthogonal axes (see Fig.~\ref{hamiltonian}). Thus the coherence  maintaining the IR could be lost thereby rendering the system to slip into a GS that is similar to the set II explaining the same behavior of cases 5, 7, and 8. Case 6 can also be simply understood from the GS perspectives, as here the GS is given by the set I which is the same GS as case 1,2, and thus the qualitative behavior and analysis go parallel as discussed above. \\
\textit{Defect:} As opposed to Model A, the one-site defect shown in Fig.~\ref{H2SQ_stat_loc}(c) and (d), we observe no significant changes in concurrence as compared to the case of no defect. However, for $J_0=-0.1, J_1=-0.05$, $J_0=-0.1, J_1=0.05$ ,$J_0=0.1,J_1=0.05$ corresponding to the cases 5,7,8 with the underlying classical GS set I, II, III, we do observe a maximum at very small fields ($\approx 0.1-0.3$). except for case 6 for which the set II (largest degeneracy and a non-IR state) show the behavior similar that is characteristic of frustration effects. The results from the significant changes in concurrence conclude that the IR states are not sensitive to concurrence rather it is the degeneracy in the system that seems to play a major role. In the next section~\ref{dynamiccoupling}, we discuss the concurrence as a function of the parameter $J_2$. 
\section{Concurrence versus intramolecular coupling}
\label{dynamiccoupling}
\subsection{Model A}
\textit{Defect free:} Here, we present the results of concurrence obtained as a function of the exchange coupling strength, $J_2$, also the intramolecular term. We vary the coupling sinusoidally as $J_2 \rightarrow J_2 \cos(\omega \gamma)$ for some $\omega$ and $\gamma$. These newly introduced constants have no physical relevance and are intended for numerical purposes such that the value of $J_2$ varies effectively from 0 to $J_2^{\text{max}}$ (here 2) within one time period, $\frac{2\pi}{\omega}$. From Fig.~\ref{dyn_modelA}, we see that the cases 1,2 and 3,4 have a concurrence value of $\sim 0.1$ and $0.037$ at $\gamma=0.7$ respectively, independent of nature of coupling constants referring to various cases. Similarly, In plot (a), the concurrence for $J_1=-0.05$ and at $K_x=0$ has maximum entanglement for all the values of $\gamma$ shown. Similarly, in plot (b), the same scenario is observed for case $J_1=-0.1$ for the same field strength.\\
\indent \textit{Defect:} Upon introducing again a site defect, we see that the concurrence gets reduced without changing the qualitative behavior. In plot (c) and (d) of Fig.~\ref{dyn_modelA}, the concurrence is plotted for one-site defect, our results show that the site defect is insensitive to exchange coupling strength as the concurrence changes only marginally compared to the case of no defect. 
\subsection{Model B}
\textit{Defect:} The concurrence here exhibits gradual increase followed by asymptotic decrease with the maximum for all the cases between the values ranging from $\gamma=0.3$ to $0.7$. The peak value of concurrence, 0.055 is comparable to the case of 12-site cluster without defect (0.12). Looking at the Figs.~\ref{H2SQ_dyn_loc}(c), we see that the case 4 corresponds to the GS with a configurational set III (see Fig.~\ref{H2SQ_dyn_loc}) and has the highest degree of entanglement compared to the other cases. Likewise, in Fig.~\ref{H2SQ_dyn_loc}(d), we see the concurrence is higher where $|J_0>|J_1|$. This is possibly due to the role of intramolecular coupling constant $J_1$ that is important dictating the amount of frustration case by case. In addition, the entanglement between qubits remain constant for some values of $\gamma$, for instance, it is 0.03527 for intermediate values of $\gamma=0.5$ to 1.0. All the cases in Fig.~\ref{H2SQ_dyn_loc}(a)-(d) have shown to have such a robust concurrence around same parameter regimes of intramolecular coupling constant. The robust nature of the concurrence could be due to the characteristic deconfined nature of the ground state. Lastly, we see that the one-site Defect on a 12-site cluster has minimal effect on the entanglement as the intramolecular coupling is increased.
\section{Summary and conclusions}
\label{summary}
We study the ground state entanglement of two transverse field models on a quasi-2D squaric lattice. Specifically, we calculate the genuine multipartite entanglement of the ground state of two different models (Model A,B) with two competing interaction strengths $J_1, J_2$ for Model A ($J_0,J_1$ for Model B). We note that  in the absence of next nearest neighbour interactions, the system is shown to host deconfined phase with exponentially diverging degeneracy in the both the models~\cite{ishizuka2011quantum,vikas2018}. We here using the analysis of 4-qubit system deduce the variation of concurrence (entanglement) along with external field and intramolecular coupling. In order to arrive at our conclusions, all the cases of the couplings as mentioned in Sec.~\ref{Gsection} were considered. In each of the cases, the couplings are different leading to different GS configurations which therefore contribute distinctly to the concurrence. Moreover, the effect of one lattice site defect on multipartite entanglement were also analysed in both the cases. We see that when the system is maximally frustrated (all the bonds are not satisfied) we see a trend where the concurrence is zero at small fields gradually increasing where a revival of entanglement takes place induced by the quantum fluctuations. Similarly, for a case with marginal frustration (where few bonds are satisfied) the concurrence behavior is intermediate of the behavior of case with GHZ states GS, i.,e the concurrence starts with non-zero value which is also not a maximum. When we introduce a defect-site in the system where the satisfied bond is now absent, we get the exact behavior as in the case of maximally frustrated scenario. This suggests that the concurrence is indeed sensitive to the amount of frustration (degeneracy) in the system. Moreover, the GHZ states also seems to be insensitive to the site-defect, the only case where the site-defect has a qualitative change in concurrence variation is the case of marginal frustration. \\ 
\indent Similar results for concurrence have been observed for Model B, however, for a site-defect case the concurrence is shown to be insensitive to it. In addition, we see that the concurrence remains constant for intermediate values of $\gamma$ ($J_2$), i.e, the intramolecular coupling strength. Such robust behavior of concurrence is believed to be because of the robust deconfined phase that is present in the corresponding parameter regime. Interestingly, the same behavior is lost as the external field strength reaches higher values ($K_x=1.5$), this is because higher values leads to the destruction of the deconfined ground state. Furthermore, the concurrence has indirect relationship with $J_1$, i.e, the intramolecular coupling, we see that when $J_1 \gg J_0$ we have a case of no frustration and no degeneracy as the system is no decoupled to give 1D interpenetrating chains and the concurrence behavior falls back to the scenario of GHZ states. From the study of ground state entanglement of two squaric acid models, we discussed various dynamics of entanglement. It is evident from the nature of entanglement under various conditions that the effects of frustration on quantum correlation are not always detrimental. As we have seen, entanglement creation along with its decay happens under frustration. In fact, the more the amount of frustration the more robust the entanglement creation and decay. 

\begin{acknowledgments}
All the computations performed were in part supported by High-performance computing facility at IOP. 
\end{acknowledgments}
\bibliography{apssamp}
\end{document}